# Magnetic precursor to the structural phase transition in $V_2O_3$

Chubin Huang,[1*] Abhishek Rakshit,[1*] Gianluca Janka,[2] Zaher Salman,[2] Andreas Suter,[2] Thomas Prokscha,[2] Benjamin A. Frandsen[3] and Yoav Kalcheim[1]

[1]Department of Material Science and Engineering, Technion – Israel Institute of Technology

[2]Laboratory for Muon Spin Spectroscopy, Paul Scherrer Institute, 5232 Villigen PSI, Switzerland

[3]Department of Physics and Astronomy, Brigham Young University, Provo, Utah 84602, USA

*both authors contributed equally to this work

**Abstract:** The coupling between structural, electronic and magnetic degrees of freedom across the metal-insulator transition in $V_2O_3$ makes it hard to determine the main driving mechanism behind the transition. Specifically, the role of magnetism has been debated and its interplay with the other transitions has not been established. To address this issue, we use a combination of muon spin relaxation/rotation, electrical transport and reciprocal space mapping which allows to correlate magnetic, electronic and structural degrees of freedom in strain-engineered $V_2O_3$ thin films. Evidence is found for a magnetic instability in the vicinity of the structural transition. This is manifested as a decrease in the antiferromagnetic moment with temperature leading to a virtual Néel transition temperature which coincides with that of the structural and electronic transitions. Moreover, we find evidence for an onset of antiferromagnetic (AF) fluctuations in the rhombohedral phase even without a structural transition to the monoclinic phase. The non-congruence of the structural and magnetic transitions increases as the transition temperature is reduced by strain. In samples where the transition is most strongly suppressed by strain, a depth-dependent magnetic state is observed. These results reveal the importance of an AF instability in the paramagnetic phase in triggering the metal-insulator transition and the crucial role of the structural transition in allowing for the formation of an ordered AF state.

The insulator to metal transition (IMT) in Mott insulators offers important opportunities for studying strong electronic interactions as well as exciting technological advancements, especially in the field of resistive switching [1–19] and sensing. [20–28] The IMT in these materials often involves coupled structural, electronic and magnetic transmutations offering rich physics and sensitivity to various parameters. [29] However, this coupling hinders the identification of the main driving mechanism behind the transition and the ability to judiciously control it. In some cases, under the influence of strain or compositional variations, these transitions may decouple, shedding light on the transition mechanism and the role of strong electronic correlations. For example, doping or stress in $VO_2$ can result in different structural insulating phases without dimerization, with implications for the role of the Peierls vs. Mott-Hubbard mechanism in establishing an insulating state. [30–32] In some Mott insulators, decoupling of the IMT and structural phase transition (SPT) [24,30,33–43] or more subtle changes to the electronic or lattice structure in the vicinity of the transition [44–46] have also been reported, suggesting either an electronic or structural origin for the transition. These are challenging experiments since they require careful comparisons between various measurement techniques probing different properties and occasionally different volumes of the sample. Clearer results are found in $RENiO_3$ (RE=Rare earth element), where the coupled structural and magnetic transition observed for RE=Pr & Nd becomes decoupled with heavier RE elements, clearly disentangling magnetic and structural ordering. [47–53]

In $V_2O_3$, one of the most canonical examples of a Mott insulator, it has recently been found that the structural and electronic transitions are robustly coupled even in the presence of large strain. [54] Also, it was recently shown in X-ray and neutron pair distribution function measurements that the structural transition starts abruptly with no signs of structural distortions or fluctuations preceding the SPT. [55] As for the role of magnetic ordering, it has been argued that orbital ordering modulates the exchange interaction and breaks translational symmetry, giving rise to the antiferromagnetic (AF) insulating phase. [56,57] Spin correlations were observed both in the monoclinic AF phase and the rhombohedral paramagnetic metal (PM) and paramagnetic insulator (PI) phases but they have different wavevectors indicating that the structural transition may strongly affect the type of magnetic correlations observed in the different phases. Indirect evidence for the importance of magnetism in the IMT have also been observed through magnetoresistance measurements which are found to change sign in the vicinity of the phase transition and grow towards the metal-to-insulator transition (MIT). [58] More recently, using muon spin relaxation/rotation (µSR) we have demonstrated the possibility of measuring magnetic phase fractions and magnetic moments across the phase transition of $V_2O_3$ thin films. [59] Building on this work, we now examine the interplay between the structural and magnetic degrees of freedom as they evolve through the phase transition in strain-engineered $V_2O_3$ thin films. Due to the strain, the films exhibit both positive and "negative" pressure regimes which decrease and increase the transition temperature with respect to bulk values, respectively. [60] We also examine samples where geometrical confinement of the film by substrate clamping changes the transition trajectory so that it follows the PM-AFI phase equilibrium line down to <2 K. [61] This concerted study reveals magnetic precursors to the SPT/IMT as the transition is approached from both the low and high temperature sides, indicating the central role of magnetism in driving the transition into the AF insulating phase of $V_2O_3$.

In this work we studied 100 nm $V_2O_3$ thin film samples which were grown using RF magnetron sputtering from a $V_2O_3$ target at a substrate temperature of ~700 C (see ref [62] for details). This procedure has been previously shown to result in heteroepitaxial growth of the $V_2O_3$ films, following the orientation of the isostructural $Al_2O_3$ substrate. [63] To produce different strain states, samples were grown on sapphire substrates with different crystallographic orientations, namely (110), (100) and (001) which will be henceforth referred to as A-cut, M-cut and C-cut samples respectively. For each orientation, growth was simultaneously performed on several substrates (with an area larger than 12mm by 18mm in total) cut from the same wafer to obtain a strong signal in µSR measurements, where the standard deviation of the muon beam spot size is ~6 millimeters. Despite having identical growth parameters, representative resistance vs temperature measurements for film grown on substrate with different orientations, show large variations in the transition characteristics, revealing the role of substrate-induced strain (Fig. 1(a)). The M-cut sample shows a resistive transition centered around 177K, and the A-cut sample shows a transition around 150 K. On the other hand, the C-cut sample exhibits a strongly suppressed resistive transition. This phenomenology has been discussed previously by some of the authors of the present work (see Fig. 1 (b)). [54,60,61] It has been found that the hexagonal ab-plane strain plays a central role in determining the transition trajectory. The M-cut films have the largest ab-plane tensile strain leading to "negative" pressure effects similar to those observed in Cr-doped $V_2O_3$. This leads to the coexistence of paramagnetic metallic and insulating phases at all temperatures above ~185 K and a high transition temperature to the AFI phase compared to bulk values of pure $V_2O_3$. The A-cut sample is less strained in the ab-plane and shows a transition temperature slightly lower than the bulk value. [60] In both A-cut and M-cut samples the ab-plane can freely expand in the out-of-plane direction which is required for the formation of the low T monoclinic-insulating phase. Contrary to this, in the C-cut samples the ab-plane is not free to expand, since the expansion is completely in the film plane and the film area is limited by the substrate. This leads to a partial transition into the low temperature phase of $V_2O_3$ and an increase in strain during the transition resulting from the regions which have transitioned into the expanded monoclinic phase. [61]

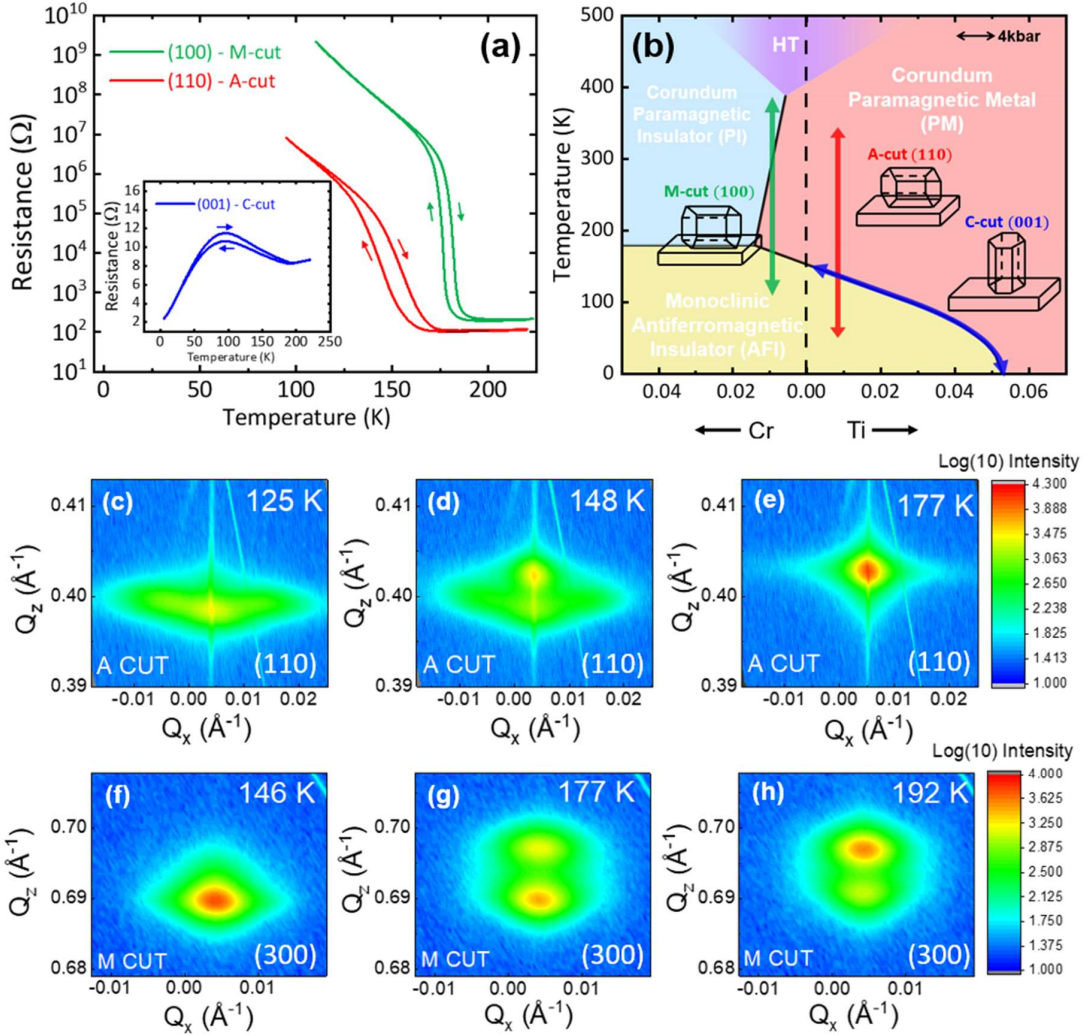

*Figure 1: **Resistive and structural transition in $V_2O_3$ films and their corresponding trajectories in the phase diagram.** (a) Resistance vs temperature curves of samples deposited on sapphire substrates cut along the three different crystallographic orientations - (100), (110) and (001), referred to as M-cut, A-cut and C-cut respectively. (b) Schematic representation of the trajectories of the phase transitions for the different samples. The M-cut sample is under strong tensile strain in the ab-plane and therefore exhibits a high transition temperature to the AFI phase and coexistence of PM and PI at high temperatures. The A-cut sample has lower ab-plane strain and therefore has a low transition temperature to the AFI phase. The C-cut sample exhibits a transition following the phase equilibrium line between the PM and AFI phases due to the clamping of the ab-plane within the film plane. See [60,61] for details. (c-e) and (f-h) Reciprocal space maps (RSM) around the (110) Bragg peak in the A-cut sample and (100) Bragg peak in the M-cut sample, respectively. RSMs were acquired during the heating cycle. Note the different temperature ranges in A-cut and M-cut films. Films were grown under identical growth conditions and differences in transition temperatures are due to strain. See SI section 6 and Fig. 3 for the extraction of structural phase fractions from RSMs.*

The structural evolution across the transition was studied using a lab-based XRD system equipped with a cryostat. Reciprocal space maps (RSMs) were acquired at various temperatures across the transition around the out-of-plane Bragg peak of each sample (see Fig. 1(c-h)). To quantify the fractions of the low and high temperature structural phases across the transition we used an RSM weighted sum fitting scheme as described in the following: RSMs acquired outside of the transition region were taken as representative of the fully monoclinic or rhombohedral phases. We then analyzed each RSM acquired at intermediate temperatures as a weighted sum of these two RSMs using a least-mean-squares method. The fitted weight of the RSM acquired at high temperature is used as a measure of the rhombohedral phase fraction. This fitting method circumvents conventional peak fitting which is less reliable due to the non-trivial RSMs observed in our films. To exactly pinpoint the completion temperature of the structural transition we have also examined differentials of the RSM, where we subtracted the RSMs acquired at consecutive temperature points. An excellent agreement is found between this approach and the least-squares sum method (see SI – section 6 for details).

The magnetic evolution of our samples was probed using μSR at the Swiss Muon Source of the Paul Scherrer Institute. To probe the magnetism of the thin films, we used the LEM (Low Energy Muons) beamline at the μE4 beamline, [64] on which a continuous beam of fully spin-polarized 4 MeV muons was decelerated by a moderator to energies of about 15 eV, then electrostatically accelerated to a targeted implantation energy ranging from 4 keV to 18 keV. This enables implantation of the muons at a controllable mean depth ranging from 20 - 80 nm in our $V_2O_3$ films. The implanted muons rapidly come to rest at an interstitial site corresponding to an electrostatic potential minimum, upon which each muon spin precesses in its local magnetic field. After a mean lifetime of 2.2 μs, the muon decays into a positron and two neutrinos, with the former emitted preferentially in the direction of the muon spin at the instant of decay. The main experimental quantity measured in a μSR experiment is the muon decay asymmetry, a dimensionless quantity computed as the normalized difference in positron counts between pairs of detectors placed opposite each other near the sample. The asymmetry is proportional to the component of the spin polarization of the muon ensemble along the direction joining the two detectors. Detailed information about the local magnetic field distribution in the sample can therefore be extracted from the time dependence of the asymmetry.[63] In a well-ordered AF state, μSR measurements in zero applied field (ZF) exhibit spontaneous oscillations with a frequency that is proportional to the intrinsic local field, and therefore provides a measure of the sublattice magnetic moment in AF regions of the sample (see SI – section 1). The paramagnetic phase fraction can be determined from data collected in the presence of a weak transverse field (wTF) of 50 G, which is small compared to the intrinsic field of the AF state in the sample. Muons landing in the paramagnetic regions of the sample result in an asymmetry component oscillating at the frequency set by the external field, while the asymmetry component from muons landing in AF regions of the sample has much faster oscillations and damping. When the asymmetry is averaged over timescales which are longer than the oscillation period of muons in the AF regions, only the oscillations of muons which landed in the paramagnetic regions contribute to the oscillating asymmetry signal. Thus, the amplitude of the slow wTF oscillations provides a measure of the paramagnetic phase fraction (see SI – section 2). We performed least-squares fits to the μSR data with the program MUSRFIT [65] to extract the ZF oscillation frequencies and wTF amplitudes. In principle, the ZF oscillation amplitude may also be used to obtain the magnetic phase fraction. However, due to the much shorter timescales over which ZF oscillations occur, the measurement statistics are significantly poorer compared to those of the wTF data. In the case of an isotropic field distribution, the non-oscillating component of the ZF data can also provide information on the magnetic phase fraction. However, in the present case of epitaxial $V_2O_3$ films the moments are strongly anisotropic and therefore a reorientation of the magnetic moments would cause changes to the non-oscillating component which is not due to a change in phase fraction. Such changes are indeed observed in the ZF data (see SI – section 1). Therefore, we focus on the wTF oscillation amplitude to measure magnetic phase fractions and ZF oscillation frequency to measure the moment magnitude.

Independent measurements of the paramagnetic phase fraction and AF moment are unique to µSR and, along with the structural characterization by XRD, provide a detailed account of the interplay between structural and magnetic properties in the $V_2O_3$ samples. One important difference between µSR and XRD is in the sampled volume. While XRD probes the entire film thickness due to the large penetration depth of the x-rays (tens of microns), the muon implantation profile depends on its energy. In a previous study it was found that in $V_2O_3$ films with a sharp transition, muons with different energies show a very similar magnetic transition as a function of temperature. [59] Therefore, for A-cut and M-cut samples which exhibit a sharp transition we selected muon energies of 14 keV which have a mean implantation depth of 58 nm, approximately in the middle of the sample. This provides an excellent representation of the magnetic transition in the middle of the sample. Conversely, in the C-cut sample it has recently been shown that the transition progresses in a strongly depth dependent manner. Therefore, in this sample we used several muon implantation energies ranging from 4-14 keV to obtain the depth-dependent magnetic transition profile (see SI - section 5).

When comparing XRD and µSR data collected on different instruments, it is also important to take into account that the nominal temperature reading in both systems may not accurately correspond to the same sample temperature. This may arise from calibration issues or differences between the sample and sensor temperatures. This is especially important in the A-cut and M-cut samples which have sharp transitions and therefore even errors of several Kelvins may lead to erroneous interpretations. To this end, Au/Ti contacts were evaporated on the sample edges and used for in-situ resistance measurement during both XRD and µSR temperature-dependent measurements. By examining the temperature shifts between the resistance vs temperature curves of each sample as measured in the different experimental configurations, an exact comparison between the two data sets is enabled with an accuracy better than 1 K (see SI section 3 for details).

Figure 2 shows a comparison of the evolution of structure and magnetism across the IMT in the M-cut and A-cut samples. Figure 2(a) shows the ZF asymmetry oscillation frequency as a function of temperature for the M-cut and A-cut samples during heating. It is important to note that the ZF oscillations are proportional to the magnitude of the moment in AF regions but are independent of the AF volume fraction. Therefore, as long as the AF volume fraction is large enough to obtain an asymmetry signal, the evolution of the AF magnetic moment can be traced regardless of the decrease in AF volume fraction. It is found that the magnitude of the AF moment in both A-cut and M-cut samples shows a very similar temperature dependence up to the temperature where they can no longer be detected (A-cut – 166 K and M-cut - 181 K). This temperature dependence can be well described by a power law of the form $f = f_0(1 - T/T_N)^\beta$ where $T_N$ represents a virtual Néel temperature corresponding to an idealized continuous magnetic phase transition, which in reality is preempted by the first-order IMT/SPT in $V_2O_3$. For both A-cut and M-cut samples good fits are obtained for a range of $\beta = 0.24 - 0.33$ with corresponding values of $T_N \approx 190 - 205\ K$. Since the IMT is not a simple Néel transition due to the first order IMT/SPT which preempts it, one might expect that $T_N$ would be unrelated to the other transitions. However, we find that $T_N$ is very close to the maximum temperature where the monoclinic phase is still observed both in the strained films of this study and in previously studied Cr-doped bulk $V_2O_3$. [66] This indicates that AF ordering plays an important role in stabilizing the insulating monoclinic phase. We note, however, that the AF moment evolution in both A-cut and M-cut samples is very similar despite their different structural/electronic transition temperatures. This indicates that strain also has a direct contribution to the stability of the various phases which is independent of the AF ordering.

We now compare the evolution of the rhombohedral and paramagnetic phases (Figure 2b and 2c). While the two transitions are close to each other, a clear difference is observed. In both samples the transition into the fully PM phase ends at a higher temperature than the structural transition. We note that the magnetic transition is wider than the structural one in both cases so that this difference is robust against errors in determination of the exact sample temperature. In the case of the A-cut sample this difference is very pronounced, where the wTF oscillations reach their maximum amplitude ~9 K above the temperature where the sample becomes fully rhombohedral, well beyond the uncertainty of our temperature calibration. These findings indicate that the rhombohedral phase of $V_2O_3$ has a magnetic instability even in the absence of an SPT.

The comparison between the ZF and wTF signals sheds light on the nature of the magnetic state which arises at the transition. Examining Fig. 2a we find that in the A-cut sample the ZF oscillations are no longer detected at 166 K whereas in the M-cut sample they persist up to 181 K. These temperatures are indicated as vertical lines in Fig. 2(b) & 2(c). Interestingly, in both A-cut and M-cut samples, at the temperature where the ZF oscillations vanish there is still a sizable non-PM phase fraction (over 30% in both cases). However, we estimate that the detection limit of ZF oscillations is less than 10% of the maximum asymmetry observed at the fully AF state. If over 30% of the sample displayed an ordered and stable AF moment it would be well within the detection limit of the ZF oscillation measurement. This indicates that a sizable fraction of the sample is not paramagnetic but also does not show the ZF oscillations expected from a stable and ordered AF state. This could have two possible explanations: 1. the presence of AF fluctuations in the PM phase: muons landing in regions which develop sub-microsecond transient AF moments. 2. A stable AF phase but with strongly disordered AF moments. Both scenarios would result in a reduced wTF oscillating asymmetry yet a non-oscillating ZF asymmetry. In the first scenario, temporal fluctuations constitute a clear magnetic instability solely based on the μSR results. In the second scenario, spatially disordered magnetic moments do not necessarily indicate a magnetic instability. However, their presence within the fully rhombohedral phase, as evidenced by the difference between PM and rhombohedral phase fractions, is a clear indication for a magnetic precursor to the structural phase transition. Further studies using x-ray linear dichroism [43] or neutron diffraction [58] could help to differentiate between these scenarios.

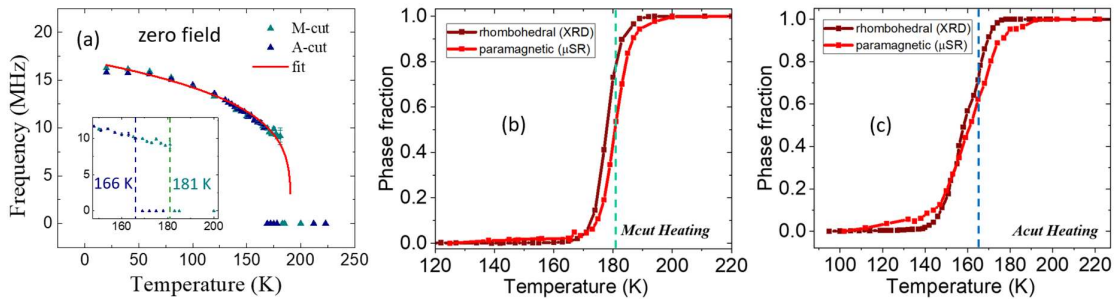

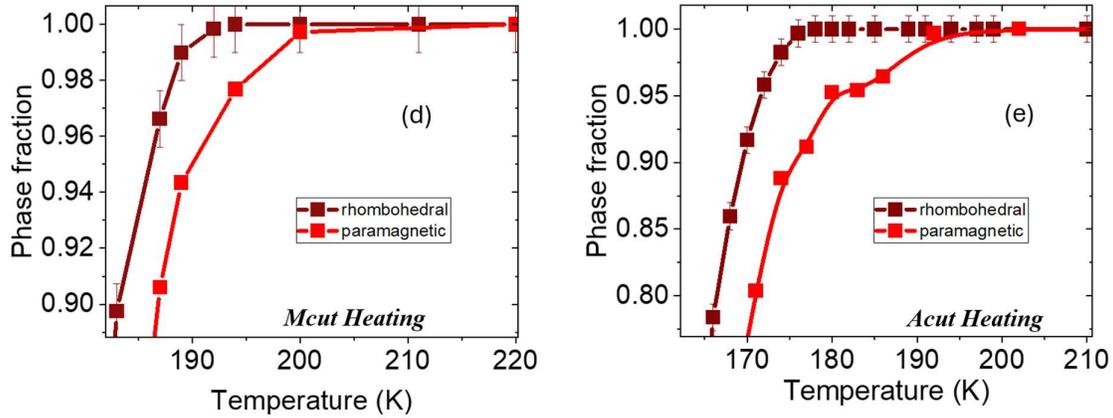

*Figure 2: Evolution of magnetic and structural transitions in strained $V_2O_3$ films. (a) zero-field muon asymmetry oscillation frequency as a function of temperature for A-cut and M-cut samples. The red line is a fit to $f = f_0(1 - T/T_N)^\beta$ - see text for details. The inset shows an enlargement of the high temperature regime and the dashed lines correspond to the temperatures above which no asymmetry oscillations could be detected. (b+c) comparison between paramagnetic phase fraction as deduced from wTF muon asymmetry oscillations and rhombohedral phase fraction obtained from XRD. Dashed lines correspond to the same temperatures as indicated to the inset to (a). (d) and (e) are enlargements of the high temperature regime of (b) and (c) respectively, focusing on the differences between the completion temperatures of structural and magnetic transitions in the A-cut and M-cut samples.*

The results from the A-cut and M-cut samples show evidence for a magnetic instability in the rhombohedral PM phase close to the MIT, manifested as a difference between the evolution of the amplitude of wTF asymmetry oscillations and the structural transition. Based on these findings, we hypothesized that this non-congruence may become more pronounced if the rhombohedral PM phase is stabilized down to lower temperatures. To test this hypothesis, we performed an additional correlated µSR/XRD study on the C-cut sample where the MIT is strongly suppressed by clamping to the substrate. A comparison between the temperature dependence of the structural and magnetic properties verifies this hypothesis. As shown in Fig. 1a (blue curve), the MIT in the C-cut sample is strongly suppressed as evidenced by the small resistive transition. This is in accordance with RSM measurements (Fig. 3) showing that the structural peak corresponding to the metallic state is mostly present even down to 95 K (lowest limit of our XRD cryostat). Analysis of the peak volume as a function of temperature, shows that 12-16%% of the C-cut sample undergoes an SPT into the monoclinic phase at 95 K (see Fig. 3 and SI - section 6). Indeed, recent TEM measurements show that C-cut films exhibit a depth-dependent transition profile upon cooling. [67] While the entire film shows an incomplete MIT, the top of the film transitions to a larger extent than the bottom as expected from the stronger clamping in the vicinity of the film-substrate interface. Since the penetration depth of x-rays is much larger than the film thickness, the RSM analysis captures the structural evolution in the entire film thickness. However, muons have energy-dependent implantation depths and therefore allow us to perform wTF measurements to probe the paramagnetic phase fraction at different depths. As the sample is cooled, the evolution of the paramagnetic phase fraction shows a substantial depth dependence. For low implantation energy (top of film), the PM phase fraction decreases substantially faster with decreasing temperature than for higher implantation energy (bottom of film) – see Fig. 3(c). To calculate the total PM fraction, we used a convolution of the muon implantation depth profile with a linearly depth-dependent PM phase fraction (see SI - section 5). This allows a comparison of the total PM and rhombohedral phase fractions as a function of temperature (Fig. 3(c) – inset). Upon cooling, the PM phase fraction is found to decrease more rapidly than the rhombohedral phase fraction. This corroborates the hypothesis that the magnetic instability of the PM phase is enhanced as the film is cooled even without the ability to undergo a structural/electronic transition due to clamping. Interestingly, we find that despite the large decrease in PM phase fraction at 20 K for muons with a mean implantation depth of 20 nm (~50%), only a small ~10% oscillating asymmetry was observed in ZF oscillations measurements at the same energy and none for higher implantation energies. This again shows that strong magnetic disorder or fluctuations are present when the system is close to the coupled structural-electronic transition. The absence of ZF asymmetry oscillations when the structural transition is suppressed points to the role of the structural transition in allowing for the formation of a stable and ordered AF state.

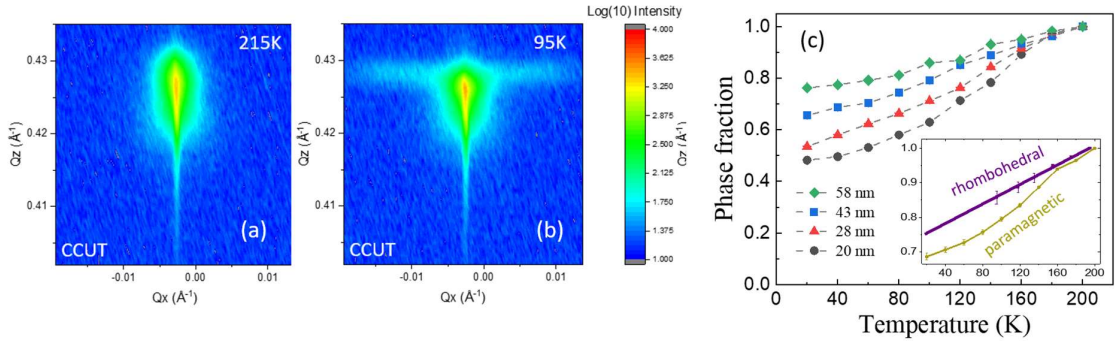

Figure 3: Structural and magnetic evolution in the (001)-oriented V2O3 film (C-cut sample). (a)+(b) reciprocal space maps (RSMs) acquired above the onset of the structural transition (a) and at 95 K (b). By tracking the intensity which shifts from the rhombohedral Bragg peak to that of the monoclinic phase as a function of temperature and dividing by the total peak intensity, the fraction of the film which undergoes a structural transition is determined. (c) Paramagnetic phase fraction as determined by wTF oscillations amplitudes for muons with different acceleration energies resulting in different mean implantation depths as indicated. The inset shows a comparison between the rhombohedral (purple) and paramagnetic (light green) phase fractions. The latter is determined from the depth dependent wTF oscillations (see text). The rhombohedral phase fraction decreases linearly with temperature while the PM phase fraction decreases more rapidly (the line is a guide to the eye).

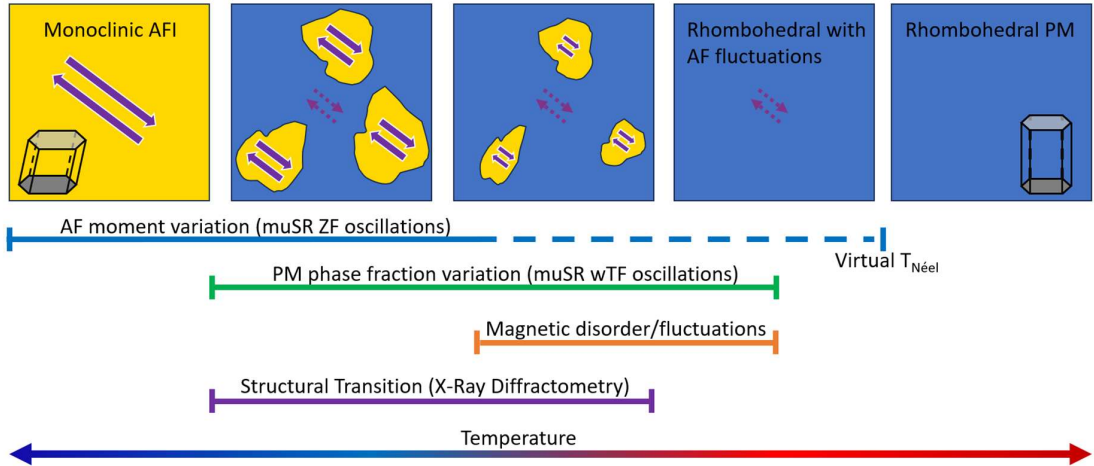

*Figure 4: Schematic representation of the main findings of this work, depicting the transition between the monoclinic antiferromagnetic insulating phase to the rhombohedral paramagnetic metallic phase (Yellow=monoclinic, Antiferromagnetic=purple arrows, blue=rhombohedral). As temperature increases, the intrinsic AF moment decreases (decreasing antiparallel arrow size). Due to magnetic disorder/fluctuations, the AF moment can no longer be detected above a certain temperature (dashed purple line) even though the PM phase fraction is significantly less than 100%. The Néel temperature corresponding to the decreasing AF moment coincides with that of the limit of stability of the monoclinic phase of $V_2O_3$, suggesting that AF ordering stabilizes the monoclinic phase. The PM phase fraction reaches 100% at temperatures above which the sample is fully rhombohedral, indicative of AF fluctuations (dashed purple arrows) in the rhombohedral phase. The non-congruence of the structural transition (blue line) and magnetic transition (green line) and the magnetic disorder/fluctuations (orange line) indicate a magnetic precursor to the structural transition.*

In conclusion, we have performed a correlated study of the structural and magnetic evolution of strongly strained films of $V_2O_3$ across the insulator-metal transition using XRD and μSR. Contrary to a scenario where magnetic ordering is a by-product of the structural/electronic transition, the temperature dependence of the AF magnetic moment indicates a virtual Néel temperature which coincides with an upper limit of stability for the monoclinic phase in $V_2O_3$. This indicates that AF ordering plays an important role in stabilizing the monoclinic phase. However, the AF moment evolves similarly in samples with the different transition temperatures, indicating that strain also plays an independent role in the transition. Furthermore, close to the transition temperature but already in the fully rhombohedral phase the films are not completely paramagnetic. Thus, in the vicinity of the transition the metallic rhombohedral phase which was presumed to be paramagnetic, exhibits a tendency towards AF ordering as well. These observations show onsets of magnetic instabilities of both the metal and insulator phases close to the transition, pointing towards the important role of magnetism in driving the transition between them. In strongly clamped $V_2O_3$ films, even though most of the structure remains rhombohedral down to very low temperatures, magnetic fluctuations become stronger, providing further evidence for a magnetic instability of the rhombohedral phase. Analysis of complementary measurement modes of μSR reveal that the metal/insulator coexistence regime is characterized either by magnetic disorder or magnetic fluctuations which are slightly faster than the muon life-time.

Acknowledgements

This work has been funded by the European Union (ERC, MOTTSWITCH, 101039986). Views and opinions expressed are however those of the author(s) only and do not necessarily reflect those of the European Union or the European Research Council Executive Agency. Neither the European Union nor the granting authority can be held responsible for them. This work is based on experiments performed at the PSI Center for Neutron and Muon Sciences CNM, 5232 Villigen PSI, Switzerland.

References:

[1] J. del Valle, J. G. Ramírez, M. J. Rozenberg, and I. K. Schuller, *Challenges in Materials and Devices for Resistive-Switching-Based Neuromorphic Computing*, J Appl Phys **124**, 211101 (2018).

[2] J. Rodríguez Contreras, H. Kohlstedt, U. Poppe, R. Waser, C. Buchal, and N. A. Pertsev, *Resistive Switching in Metal–Ferroelectric–Metal Junctions*, Appl Phys Lett **83**, 4595 (2003).

[3] E. Janod et al., *Resistive Switching in Mott Insulators and Correlated Systems*, Adv Funct Mater **25**, 6287 (2015).

[4] Y. Kalcheim, A. Camjayi, J. del Valle, P. Salev, M. Rozenberg, and I. K. Schuller, *Non-Thermal Resistive Switching in Mott Insulator Nanowires*, Nat Commun **11**, 2985 (2020).

[5] P. Stoliar, M. Rozenberg, E. Janod, B. Corraze, J. Tranchant, and L. Cario, *Nonthermal and Purely Electronic Resistive Switching in a Mott Memory*, Phys Rev B **90**, 45146 (2014).

[6] P. Salev, J. del Valle, Y. Kalcheim, and I. K. Schuller, *Giant Nonvolatile Resistive Switching in a Mott Oxide and Ferroelectric Hybrid*, Proceedings of the National Academy of Sciences **116**, 8798 LP (2019).

[7] G. Mazza, A. Amaricci, M. Capone, and M. Fabrizio, *Field-Driven Mott Gap Collapse and Resistive Switch in Correlated Insulators*, Phys Rev Lett **117**, 176401 (2016).

[8] P. Stoliar, L. Cario, E. Janod, B. Corraze, C. Guillot-Deudon, S. Salmon-Bourmand, V. Guiot, J. Tranchant, and M. Rozenberg, *Universal Electric-Field-Driven Resistive Transition in Narrow-Gap Mott Insulators*, Advanced Materials **25**, 3222 (2013).

[9] A. G. Shabalin, J. del Valle, N. Hua, M. J. Cherukara, M. V Holt, I. K. Schuller, and O. G. Shpyrko, *Nanoscale Imaging and Control of Volatile and Non-Volatile Resistive Switching in VO2*, Small **16**, 2005439 (2020).

[10] P. Stoliar et al., *Resistive Switching Induced by Electric Pulses in a Single-Component Molecular Mott Insulator*, The Journal of Physical Chemistry C **119**, 2983 (2015).

[11] J. del Valle, P. Salev, Y. Kalcheim, and I. K. Schuller, *A Caloritronics-Based Mott Neuristor*, Sci Rep **10**, 4292 (2020).

[12] J. del Valle, Y. Kalcheim, J. Trastoy, A. Charnukha, D. N. Basov, and I. K. Schuller, *Electrically Induced Multiple Metal-Insulator Transitions in Oxide Nanodevices*, Phys Rev Appl **8**, 54041 (2017).

[13] J. del Valle, P. Salev, S. Gariglio, Y. Kalcheim, I. K. Schuller, and J.-M. Triscone, *Generation of Tunable Stochastic Sequences Using the Insulator–Metal Transition*, Nano Lett **22**, 1251 (2022).

[14] N. J. McLaughlin, Y. Kalcheim, A. Suceava, H. L. Wang, I. K. Schuller, and C. H. R. Du, *Quantum Sensing of Insulator-to-Metal Transitions in a Mott Insulator*, Adv Quantum Technol **n/a**, 2000142 (2021).

[15] M. Lange, S. Guénon, Y. Kalcheim, T. Luibrand, N. M. Vargas, D. Schwebius, R. Kleiner, I. K. Schuller, and D. Koelle, *Imaging of Electrothermal Filament Formation in a Mott Insulator*, Phys Rev Appl **16**, 54027 (2021).

[16] S. Oh, Y. Shi, J. del Valle, P. Salev, Y. Lu, Z. Huang, Y. Kalcheim, I. K. Schuller, and D. Kuzum, *Energy-Efficient Mott Activation Neuron for Full-Hardware Implementation of Neural Networks*, Nat Nanotechnol (2021).

[17] E. Qiu et al., *Stochasticity in the Synchronization of Strongly Coupled Spiking Oscillators*, Appl Phys Lett **122**, 94105 (2023).

[18] J. del Valle et al., *Subthreshold Firing in Mott Nanodevices*, Nature **569**, 388 (2019).

[19] J. del Valle et al., *Spatiotemporal Characterization of the Field-Induced Insulator-to-Metal Transition*, Science (1979) **373**, 907 (2021).

[20] M. Liu, J. Hoffman, J. Wang, J. Zhang, B. Nelson-Cheeseman, and A. Bhattacharya, *Non-Volatile Ferroelastic Switching of the Verwey Transition and Resistivity of Epitaxial Fe3O4/PMN-PT (011)*, Sci Rep **3**, 1876 (2013).

[21] E. Strelcov, Y. Lilach, and A. Kolmakov, *Gas Sensor Based on Metal–Insulator Transition in VO2 Nanowire Thermistor*, Nano Lett **9**, 2322 (2009).

[22] J. M. Baik, M. H. Kim, C. Larson, C. T. Yavuz, G. D. Stucky, A. M. Wodtke, and M. Moskovits, *Pd-Sensitized Single Vanadium Oxide Nanowires: Highly Responsive Hydrogen Sensing Based on the Metal–Insulator Transition*, Nano Lett **9**, 3980 (2009).

[23] B. Hu, Y. Ding, W. Chen, D. Kulkarni, Y. Shen, V. V Tsukruk, and Z. L. Wang, *External-Strain Induced Insulating Phase Transition in VO2 Nanobeam and Its Application as Flexible Strain Sensor*, Advanced Materials **22**, 5134 (2010).

[24] B.-J. Kim, Y. W. Lee, B.-G. Chae, S. J. Yun, S.-Y. Oh, H.-T. Kim, and Y.-S. Lim, *Temperature Dependence of the First-Order Metal-Insulator Transition in VO2 and Programmable Critical Temperature Sensor*, Appl Phys Lett **90**, 23515 (2007).

[25] C. Chen, X. Yi, X. Zhao, and B. Xiong, *Characterizations of VO2-Based Uncooled Microbolometer Linear Array*, Sens Actuators A Phys **90**, 212 (2001).

[26] H. T. Zhang et al., *Organismic Materials for beyond von Neumann Machines*, Applied Physics Reviews.

[27] H. Navarro et al., *A Hybrid Optoelectronic Mott Insulator*, Appl Phys Lett **118**, 141901 (2021).

[28] N. A. Butakov et al., *Broadband Electrically Tunable Dielectric Resonators Using Metal–Insulator Transitions*, ACS Photonics **5**, 4056 (2018).

[29] M. Imada, A. Fujimori, and Y. Tokura, *Metal-Insulator Transitions*, Rev Mod Phys **70**, 1039 (1998).

[30] B. S. Mun et al., *Observation of Insulating–Insulating Monoclinic Structural Transition in Macro-Sized VO2 Single Crystals*, Physica Status Solidi (RRL) – Rapid Research Letters **5**, 107 (2011).

[31] J. Cao et al., *Extended Mapping and Exploration of the Vanadium Dioxide Stress-Temperature Phase Diagram*, Nano Lett **10**, 2667 (2010).

[32] J. Cao et al., *Strain Engineering and One-Dimensional Organization of Metal–Insulator Domains in Single-Crystal Vanadium Dioxide Beams*, Nat Nanotechnol **4**, 732 (2009).

[33] S. Kumar, J. P. Strachan, A. L. D. Kilcoyne, T. Tyliszczak, M. D. Pickett, C. Santori, G. Gibson, and R. S. Williams, *The Phase Transition in VO2 Probed Using X-Ray, Visible and Infrared Radiations*, Appl Phys Lett **108**, 73102 (2016).

[34] Y.-G. Jeong, S. Han, J. Rhie, J.-S. Kyoung, J.-W. Choi, N. Park, S. Hong, B.-J. Kim, H.-T. Kim, and D.-S. Kim, *A Vanadium Dioxide Metamaterial Disengaged from Insulator-to-Metal Transition*, Nano Lett **15**, 6318 (2015).

[35] Z. Tao, T.-R. T. Han, S. D. Mahanti, P. M. Duxbury, F. Yuan, C.-Y. Ruan, K. Wang, and J. Wu, *Decoupling of Structural and Electronic Phase Transitions in $VO_2$*, Phys Rev Lett **109**, 166406 (2012).

[36] M. K. Liu et al., *Anisotropic Electronic State via Spontaneous Phase Separation in Strained Vanadium Dioxide Films*, Phys Rev Lett **111**, 96602 (2013).

[37] A. S. McLeod et al., *Nanotextured Phase Coexistence in the Correlated Insulator $V_2O_3$*, Nat Phys **13**, 80 (2016).

[38] B.-J. Kim, Y. W. Lee, S. Choi, J.-W. Lim, S. J. Yun, H.-T. Kim, T.-J. Shin, and H.-S. Yun, *Micrometer X-Ray Diffraction Study of $VO_2$ Films: Separation between Metal-Insulator Transition and Structural Phase Transition*, Phys Rev B **77**, 235401 (2008).

[39] J. Laverock, S. Kittiwatanakul, A. A. Zakharov, Y. R. Niu, B. Chen, S. A. Wolf, J. W. Lu, and K. E. Smith, *Direct Observation of Decoupled Structural and Electronic Transitions and an Ambient Pressure Monocliniclike Metallic Phase of $VO_2$*, Phys Rev Lett **113**, 216402 (2014).

[40] M. Yang et al., *Suppression of Structural Phase Transition in VO2 by Epitaxial Strain in Vicinity of Metal-Insulator Transition*, Sci Rep **6**, 23119 (2016).

[41] S. Kittiwatanakul, S. A. Wolf, and J. Lu, *Large Epitaxial Bi-Axial Strain Induces a Mott-like Phase Transition in VO2*, Appl Phys Lett **105**, 73112 (2014).

[42] J. Nag, R. F. Haglund, E. Andrew Payzant, and K. L. More, *Non-Congruence of Thermally Driven Structural and Electronic Transitions in VO2*, J Appl Phys **112**, 103532 (2012).

[43] A. Ronchi et al., *Nanoscale Self-Organization and Metastable Non-Thermal Metallicity in Mott Insulators*, Nat Commun **13**, 3730 (2022).

[44] J. Kündel, P. Pontiller, C. Müller, G. Obermeier, Z. Liu, A. A. Nateprov, A. Hörner, A. Wixforth, S. Horn, and R. Tidecks, *Direct Observation of the Lattice Precursor of the Metal-to-Insulator Transition in V2O3 Thin Films by Surface Acoustic Waves*, Appl Phys Lett **102**, 101904 (2013).

[45] S. S. Majid, D. K. Shukla, F. Rahman, K. Gautam, R. J. Choudhary, V. G. Sathe, and D. M. Phase, *Stabilization of Metallic Phase in V2O3 Thin Film*, Appl Phys Lett **110**, 173101 (2017).

[46] M. M. Qazilbash et al., *Mott Transition in VO2 Revealed by Infrared Spectroscopy and Nano-Imaging*, Science (1979) **318**, 1750 (2007).

[47] S. Catalano, M. Gibert, J. Fowlie, J. Íñiguez, J.-M. Triscone, and J. Kreisel, *Rare-Earth Nickelates R NiO₃ : Thin Films and Heterostructures*, Reports on Progress in Physics **81**, 046501 (2018).

[48] S. Lee, R. Chen, and L. Balents, *Landau Theory of Charge and Spin Ordering in the Nickelates*, Phys Rev Lett **106**, 16405 (2011).

[49] M. Medarde, P. Lacorre, K. Conder, F. Fauth, and A. Furrer, *Giant $^{16}O\text{-}^{18}O$ Isotope Effect on the Metal-Insulator Transition of $\mathit{R}\mathrm{NiO}_3$ Perovskites ( $\mathit{R} = $ Rare Earth)*, Phys Rev Lett **80**, 2397 (1998).

[50] I. Vobornik, L. Perfetti, M. Zacchigna, M. Grioni, G. Margaritondo, J. Mesot, M. Medarde, and P. Lacorre, *Electronic-Structure Evolution through the Metal-Insulator Transition in $R\mathrm{NiO}_3$*, Phys Rev B **60**, R8426 (1999).

[51] J.-S. Zhou and J. B. Goodenough, *Chemical Bonding and Electronic Structure of $R\mathrm{NiO}_3$ $(R=\mathrm{rare}$ Earth)*, Phys Rev B **69**, 153105 (2004).

[52] M. L. Medarde, *Structural, Magnetic and Electronic Properties of Perovskites (R = Rare Earth)*, Journal of Physics: Condensed Matter **9**, 1679 (1997).

[53] G. Catalan, *Progress in Perovskite Nickelate Research*, Phase Transitions **81**, 729 (2008).

[54] Y. Kalcheim, N. Butakov, N. M. Vargas, M.-H. Lee, J. del Valle, J. Trastoy, P. Salev, J. Schuller, and I. K. Schuller, *Robust Coupling between Structural and Electronic Transitions in a Mott Material*, Phys Rev Lett **122**, 57601 (2019).

[55] E. R. A. Fletcher, K. Higashi, Y. Kalcheim, H. Kageyama, and B. A. Frandsen, *Uniform Structural Phase Transition in $\mathrm{V}_2\mathrm{O}_3$ without Short-Range Distortions of the Local Structure*, Phys Rev B **104**, 184115 (2021).

[56] W. Bao, C. Broholm, G. Aeppli, P. Dai, J. M. Honig, and P. Metcalf, *Dramatic Switching of Magnetic Exchange in a Classic Transition Metal Oxide: Evidence for Orbital Ordering*, Phys Rev Lett **78**, 507 (1997).

[57] J. C. Leiner et al., *Frustrated Magnetism in Mott Insulating $(\mathrm{V}_{1-x}\mathrm{Cr}_x)_2\mathrm{O}_3$*, Phys Rev X **9**, 11035 (2019).

[58] J. Trastoy et al., *Magnetic Field Frustration of the Metal-Insulator Transition in $\mathrm{V}_2\mathrm{O}_3$*, Phys Rev B **101**, 245109 (2020).

[59] B. A. Frandsen et al., *Intertwined Magnetic, Structural, and Electronic Transitions in $V_2O_3$*, Phys Rev B **100**, 235136 (2019).

[60] E. Barazani, D. Das, C. Huang, A. Rakshit, C. Saguy, P. Salev, J. del Valle, M. C. Toroker, I. K. Schuller, and Y. Kalcheim, *Positive and Negative Pressure Regimes in Anisotropically Strained V2O3 Films*, Adv Funct Mater **n/a**, 2211801 (2023).

[61] Y. Kalcheim, C. Adda, P. Salev, M.-H. Lee, N. Ghazikhanian, N. M. Vargas, J. del Valle, and I. K. Schuller, *Structural Manipulation of Phase Transitions by Self-Induced Strain in Geometrically Confined Thin Films*, Adv Funct Mater **n/a**, 2005939 (2020).

[62] J. Trastoy, Y. Kalcheim, J. del Valle, I. Valmianski, and I. K. Schuller, *Enhanced Metal–Insulator Transition in $V_2O_3$ by Thermal Quenching after Growth*, J Mater Sci **53**, 9131 (2018).

[63] I. Valmianski, J. G. Ramirez, C. Urban, X. Batlle, and I. K. Schuller, *Deviation from Bulk in the Pressure-Temperature Phase Diagram of $V_2O_3$ Thin Films*, Phys Rev B **95**, 155132 (2017).

[64] T. Prokscha, E. Morenzoni, K. Deiters, F. Foroughi, D. George, R. Kobler, A. Suter, and V. Vrankovic, *The New ME4 Beam at PSI: A Hybrid-Type Large Acceptance Channel for the Generation of a High Intensity Surface-Muon Beam*, Nucl Instrum Methods Phys Res A **595**, 317 (2008).

[65] A. Suter and B. M. Wojek, *Musrfit: A Free Platform-Independent Framework for MSR Data Analysis*, Phys Procedia **30**, 69 (2012).

[66] B. A. Frandsen et al., *Volume-Wise Destruction of the Antiferromagnetic Mott Insulating State through Quantum Tuning*, Nat Commun **7**, (2016).

[67] A. Pofelski, S. Valencia, Y. Kalcheim, P. Salev, A. Rivera, C. Huang, M. A. Mawass, F. Kronast, I. K. Schuller, and Y. Zhu, *Domain Nucleation across the Metal-Insulator Transition of Self-Strained V2O3 Films*, ArXiv Preprint ArXiv:2312.09051 (2023).

# Supplementary Information - Magnetic precursor to the structural phase transition in $V_2O_3$

Chubin Huang,[1*] Abhishek Rakshit,[1*] Gianluca Janka,[2] Zaher Salman,[2] Andreas Suter,[2] Thomas Prokscha,[2] Benjamin A. Frandsen[3] and Yoav Kalcheim[1]

[1]Department of Material Science and Engineering, Technion – Israel Institute of Technology

[2]Laboratory for Muon Spin Spectroscopy, Paul Scherrer Institute, 5232 Villigen PSI, Switzerland

[3]Department of Physics and Astronomy, Brigham Young University, Provo, Utah 84602, USA

*both authors contributed equally to this work

### Section 1: µSR measurements in zero field (ZF) configuration

ZF field µSR measurements enable to probe the intrinsic antiferromagnetic moment in the sample. Upon implantation in the sample, the muon spin precesses about the local intrinsic field and then decays into two neutrinos and a positron. The normalized difference in positron counts measured across pairs of oppositely placed detectors near the sample is known as µSR asymmetry. If long-ranged magnetically ordered moments are present in portions of the sample, the asymmetry will oscillate as a function of time. These oscillations in the µSR asymmetry with time are referred to as ZF µSR time spectra. Representative ZF µSR time spectra of A-cut sample are depicted in **Figure S1**. Least-squares fits to the ZF time spectra are done to extract the frequency of the µSR asymmetry oscillations. This extracted frequency at a given temperature is directly proportional to the local antiferromagnetic moment at that temperature. The ZF asymmetry oscillation frequency as a function of temperature is shown in Figure 2 (a) in the main text. Above the transition temperature to the paramagnetic state, the oscillations in asymmetry relax very slowly with time, and least square fitting can no longer be done to extract an oscillating signal. Thus, at these temperatures the frequency is indicated as zero (see Figure 2 (a)).

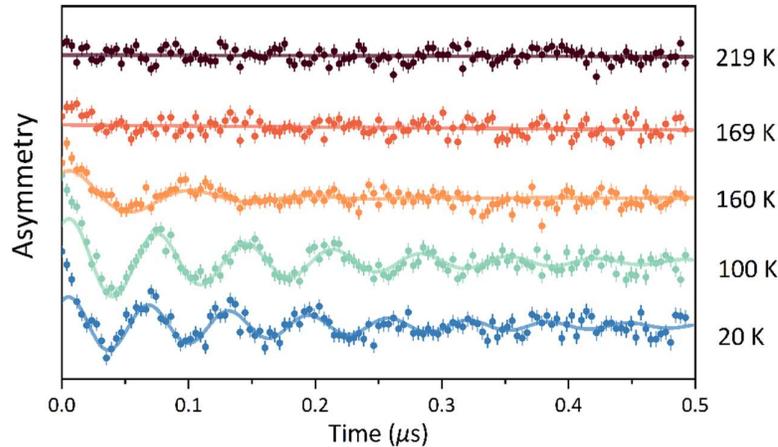

*Figure S1:* Representative zero field µSR time spectra for the A-cut sample showing oscillations in the asymmetry. The measurement was done at different temperatures during a heating cycle. Oscillation frequencies are extracted from least-squares fitting of these spectra as shown in Fig. 2 of the main text.

It is worth noting that the ZF data exhibit a fast decay in the first 10 ns of the measurement. The most probable cause for this is that in the low energy muons setup, muons arrive at the sample with a time-of-flight distribution of ~10 ns. So, at a time of 10 ns from the start of the measurement, there are muons contributing to the spectra, which are either still in flight (and thus have 100% polarization) or already stopped in the sample (with decaying polarization). Therefore, instead of measuring the pure time-dependence of the muon polarization, the initial part of the spectra is complicated by the TOF distribution of the muons. This contribution cannot be separated usually, so we cannot get useful information about the muon behavior in the sample within the first few tens of nanoseconds.

We note that in addition to the oscillating component there is also a non-oscillating decaying component which is related to muons landing in either non-magnetic regions or regions where the magnetic field is close to the direction of the muon polarization. At very low temperatures this component has a 10% contribution to the total asymmetry. However, as the temperature rises above 60 K, the magnitude of this component increases slowly until 160 K where a sharp upturn occurs until 200 K. While the sharp upturn is to be expected due to the abrupt AF-to-PM transition, the initial increase between 60 K – 160 K is not expected. We note that in this low temperature regime no changes are observed in wTF asymmetry signal (see SI section 2). Therefore, this increase is not related to an increase in the non-magnetic phase fraction. Instead, it could be attributed to a reorientation of the magnetic moments which increase the fraction of muons with polarization close to the direction of the internal magnetic field, and therefore do not precess. This finding points to the advantage of using wTF data to extract the PM phase fraction over ZF data, as explained in SI section 2.

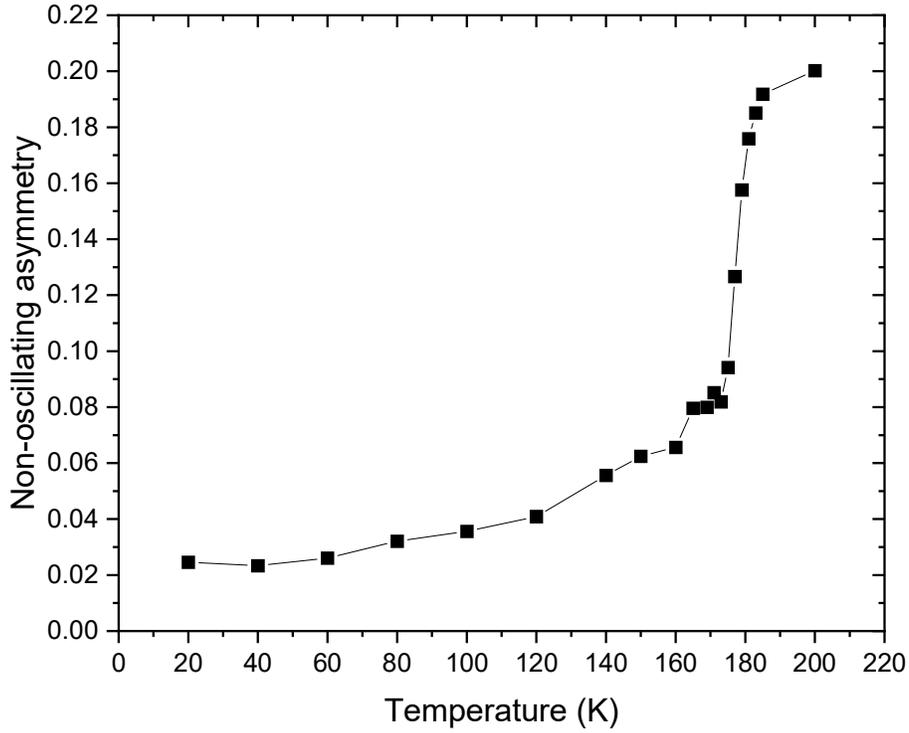

*Figure S2*: non-oscillating component of the ZF muon spectra in the M-cut sample upon warming, showing a gradual increase between 60-160 K which is not accompanied by a magnetic transition. The sharp upturn at 160 K is due to the AFI-PM transition.

### *Section 2: µSR measurements in weak transverse field (wTF) configuration*

In this configuration, a weak external magnetic field (~ 50 G) is applied out-of-plane and transverse to the initial muon spin polarization and used for probing the paramagnetic phase fraction in our $V_2O_3$ films. After implantation into the sample, the muons start precessing around the local field at the implantation site. In the paramagnetic regions of the sample the muon undergoes slow precession around the weak external magnetic field only. In contrast to this, the muons implanted in magnetically ordered regions (antiferromagnetic sites in our case) undergo fast precession about the vector sum of the local intrinsic field and the weak external field. Therefore, muons landing in these magnetically ordered regions do not contribute to the slowly oscillating signal. The amplitude of the slow oscillations is thus proportional to the volume fraction of paramagnetic phase in the $V_2O_3$ films. Representative *w*TF spectra acquired on the A-Cut sample are shown in **Figure S3**. The temperature-dependent amplitude *a(T)* of the slow oscillations arising from muons in paramagnetic regions is compared to the measured initial total asymmetry above the transition $a_{max}$. The paramagnetic phase fraction is thus derived as $\frac{[a(T)-a_{min}]}{a_{max}-a_{min}}$. In this expression, $a_{min}$ is the minimum oscillating amplitude in the fully antiferromagnetic state which is attributed to the muons landing outside of the $V_2O_3$ film. Variation of the calculated paramagnetic phase fraction with temperature for A-cut and M-cut samples is depicted in Figure 2(b)-(c) of the main manuscript.

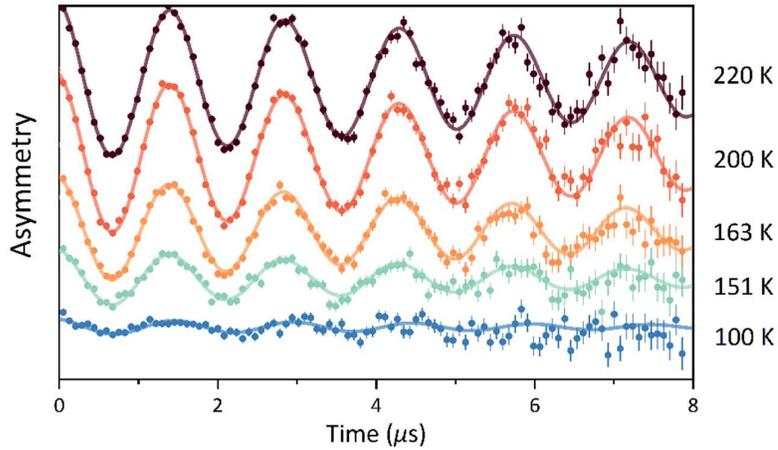

**Figure S3:** *Representative weak transverse field (~50 G) µSR spectra collected from A-cut sample at different temperatures in a heating cycle.*

### Section 3: Temperature calibration across different measurements using resistance vs. temperature as an internal thermometer

Resistance vs. temperature (*R–T*) measurements were performed in a cryogenic probe station to probe the insulator–to–metal transition (IMT) of A-cut, M-cut and C-cut samples as shown in Figure 1 (a) in the main text. During the µSR and XRD measurements, it is necessary to ensure that the measured temperatures match each other to eliminate possible offsets due to differences in calibration or thermal gradients between the sensors and the samples. This was enabled by performing *in-situ* resistance measurements during the temperature-dependent XRD and µSR studies and comparing them to a reference RT curve acquired in a cryogenic probe station. For instance, it was found that the XRD and µSR *in-situ* resistance heating curve is shifted by 2 K to lower temperature values for the A-cut sample compared to the RT measured in the probe-station (**Figure S4**). Thus, these curves were shifted by 2 K to higher temperature values to correspond exactly to the probe station RT curve. Similarly, the heating cycles of paramagnetic phase fraction and rhombohedral phase fraction data (**Figure 2(c)** main text) were shifted by 2 K to higher temperature values. We note that the *relative* correction between the µSR and XRD data were smaller than 1 K in all cases. This temperature calibration ensures accurate depiction of structural and magnetic transitions as a function of temperature for our samples.

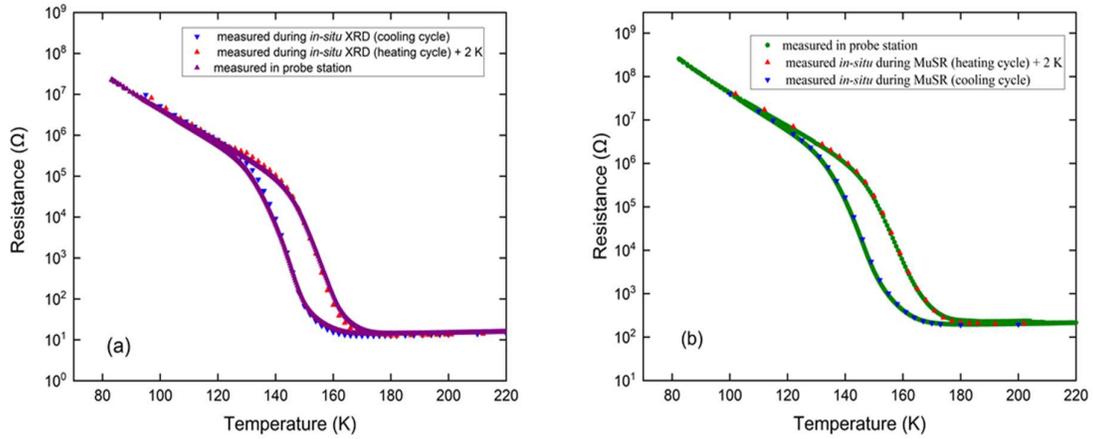

*Figure S4:* Resistance vs. temperature curves for the **A-cut sample** measured (a) in-situ during XRD and in a cryogenic probe station and (b) in-situ during µSR and in the probe station. The heating curves of the resistance data measured in-situ during XRD and µSR were shifted by 2 K, in order to match the probe station data. This shift was done to ensure that the temperature reading in all three systems correspond to the same sample temperature.

Moreover, in-situ resistance measurements allow us to ensure that the state of the sample is stable before and during the measurement as shown in **Figure S5**.

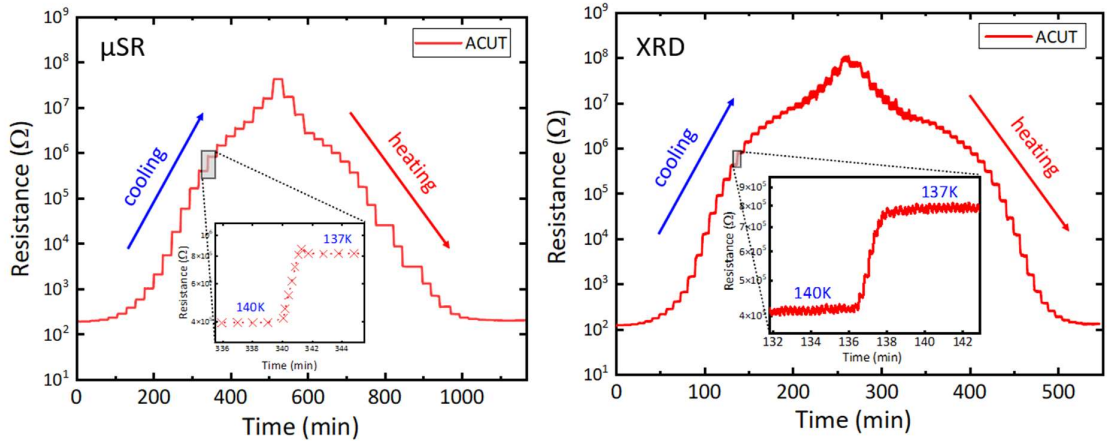

*Figure S5:* Resistance vs. time curves measured for the A-cut sample during XRD (a) and muSR measurements (b). For each measurement point, the temperature was stabilized for 3 minutes before starting the data acquisition. This ensures that the sample is in equilibrium conditions at each measurement point.

We also ensured that the temperature ramp rates we have used (2K/min) are slow enough to create quasistatic conditions during the temperature ramp. As shown **in Figure S6** there is negligible difference between R(T) curves acquired at ramp rates of 2K/min and 5K/min. Since resistance provides a highly sensitive indication for the state of the system, it is found that 2K/min is well within the quasistatic conditions.

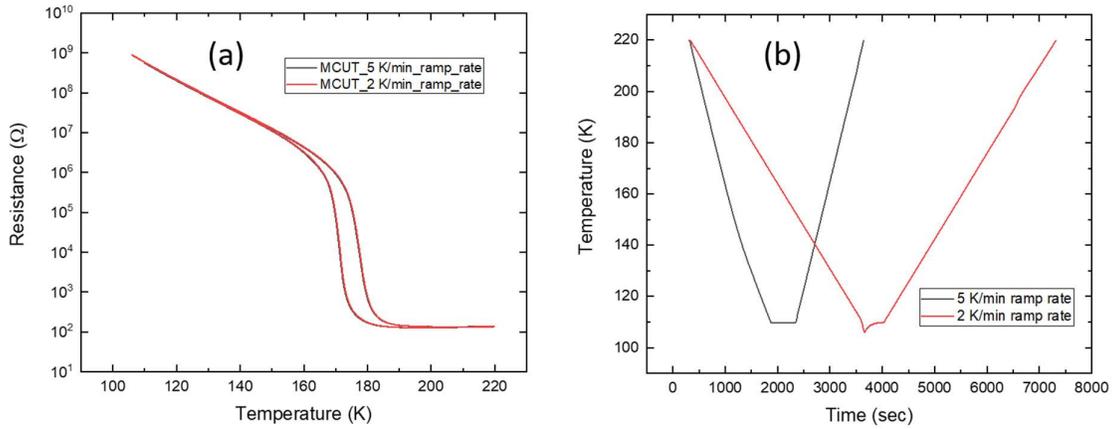

*Figure S6:* Absence of ramp rate dependence on R(T) curve for an M-cut $V_2O_3$ film. The R(T) curves are shown in (a) while (b) shows the temperature ramp profile used to acquire the data in (a).

### Section 4: Difference between magnetic and structural phase fractions during cooling:

Similarly to results obtained during heating (Fig. 2 (b&c) in the main text), a difference between magnetic and structural phase fraction evolution is observed also upon cooling.

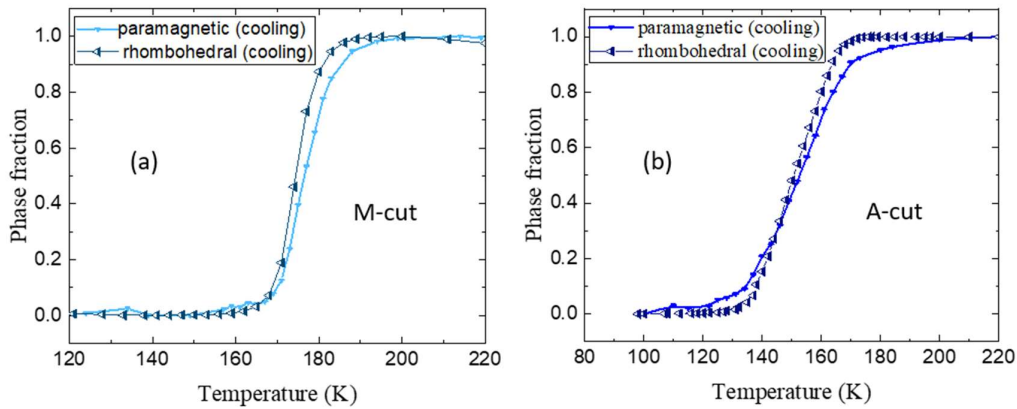

*Figure S7:* Dependence of magnetic and structural phase fraction during cooling for M-cut (a) and A-cut $V_2O_3$ films. In both cases, a clear difference is observed in the high temperature regime of the transition, indicating a magnetic instability which precedes the structural transition (see main text for details.

### Section 5: Depth–dependent paramagnetism study for C-cut sample by tuning the muon incident energies in wTF μSR measurements

The paramagnetic phase fraction vs. temperature data for C-cut sample at different muon incident energies is shown in Figure 3(c) of the main text. The mean implantation depths of the muons are directly proportional to their incident energies. As shown in **Figure S8**, the peak of the μSR energy distribution shifts to increasing thickness/depths with increasing muon incident energies. Thus, in a wTF configuration, the depth-dependance of the paramagnetism of the C-cut sample could be probed by increasing the energy of incident muons. To determine the PM phase fraction we performed a convolution of the implantation profile $P(z)$ as a function of depth with a linearly depth dependent PM profile of the form $az + b$: $\int_0^t P(z)(az + b)dz$, where t is the thickness of the sample (100 nm). For each temperature the parameters a and b were optimized using least squares to fit the measured muon-energy dependent PM values as shown in Fig. 3(c) of the main text. The total PM phase fraction of the film is then $PM_{tot}(T) = 1 - \frac{(1-b(T))^2}{2a(T)t}$. The values of $PM_{tot}(T)$ are shown in the yellow curve of the inset to Fig. 3(c).

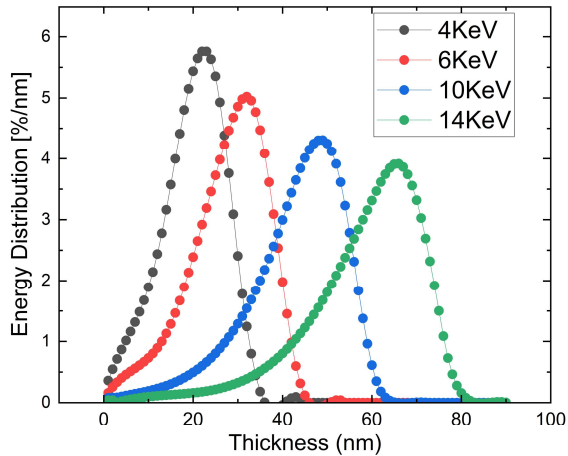

**Figure S8:** Energy distribution of muons with film thickness at different muon incident energies for the C-cut sample. Mean implantation depths for 4keV, 6keV, 10 keV and 14keV are 20 nm, 28 nm, 43 nm and 58 nm, respectively.

### Section 6: Reciprocal Space Mapping (RSM) analysis and extraction of phase fractions

To quantify the fractions of the low and high temperature structural phases across the structural phase transitions in A-cut and M-cut samples, we used an RSM weighted sum fitting scheme. RSMs acquired outside the transition region were taken as representative of the fully monoclinic or rhombohedral phases. The RSM at 220 K represents the fully rhombohedral phase whereas the RSM at 94 K represents the fully monoclinic phase. We then analyzed each RSM acquired at intermediate temperatures as a weighted sum of these two RSMs using a least-mean-squares method:

$$RSM(T) = \alpha(T) \cdot RSM_{highT} + \beta(T) \cdot RSM_{lowT}$$

*where $\alpha$ and $\beta$ are fitting parameters, and the phase fraction of the rhombohedral phase as a function of T is* $\frac{\alpha(T)}{\alpha(T)+\beta(T)}$. The fitted weight of the RSM acquired at high temperature is used as a measure of the rhombohedral phase fraction. This fitting method circumvents conventional peak fitting which is less reliable due to the non-trivial RSMs observed in our films. This also implicitly takes into account the structure factors of the low and high temperature Bragg peaks since they determine the intensity of the RSMs at high and low temperatures.

Quantifying the evolution of the SPT in C-cut samples requires a different approach to that used in the A-cut and M-cut samples. Since the transition in C-cut samples remains incomplete at all temperatures, the low temperature RSM does not represent the fully monoclinic phase. Therefore, we cannot treat RSMs as a sum of two end-states as in the A-cut and M-cut samples. We thus calculate the fraction of the sample which undergoes the SPT using the differentials of the RSMs as a function of temperature relative to the high-temperature state at 195 K. As an example, **Figure S9** shows the difference between the RSMs acquired at two extreme temperatures 195 K and 95 K. As the temperature decreases, there is a noticeable shift in Bragg peak intensity from an area in reciprocal space denoted as A to two wing-shaped monoclinic phases, B and C. We note that in bulk $V_2O_3$ the rhombohedral c-axis decreases in length during the transition from the high-T rhombohedral phase to the low-T monoclinic phase. Here, the opposite effect is observed, i.e. an increase of the c-axis length of the monoclinic phase relative to the rhombohedral phase. This is attributed to the increasing in-plane stress which is caused by the expansion of the a-axis due to the transition to the monoclinic phase. As the a-axis expands due to the monoclinic distortion, the increasing in-plane stress results in an out-of-plane expansion due to the Poisson effect, thereby elongating the c-axis. This increasing stress also acts on the rhombohedral regions causing a spectral shift from A towards D. This phenomenon is not observed in the other films where spectral weight shifts between peaks and not within the same peak. We note that due to symmetry considerations, an additional monoclinic peak is expected at $Q_x=0$. However, due to the overlap with the rhombohedral peak it cannot be resolved. Therefore, to obtain an upper bound on the fraction of the sample which undergoes the monoclinic distortion we consider the total intensity gained in regions B and C of reciprocal space (excluding region D which is part of the rhombohedral peak) and multiply it by 1.5 to account for the third monoclinic peak. These values are derived from the temperature dependent RSMs and normalized against the total peak intensity observed at 195K, as shown in Fig 3c of the main text (see also table S10). We note that at 95 K the total intensity in the RSM drops by only 0.5 % compared to 195 K, which suggests that the structure factor of the monoclinic peak may be up to 5-6% smaller than that of the rhombohedral peak. The monoclinic peak intensity increases to ~12% of the total intensity at 95 K. We can also estimate directly the change in the rhombohedral peak which yields a similar a reduction of ~16% at 95 K. The rhombohedral phase fraction corresponding to these two analyses methods appear as the error bars in the revised Fig. 3 of the main text. In both analysis scenarios we assume that this peak overlaps entirely with the rhombohedral one, which probably overestimates the monoclinic phase fraction (and therefore underestimates the rhombohedral one). As shown in Fig. 3, even under this assumption the rhombohedral phase fraction is consistently higher than the paramagnetic phase fraction at all temperatures. This observation is qualitatively the same as that found in the A-cut and M-cut samples but occurs over a much broader temperature range due to the substrate clamping effect.

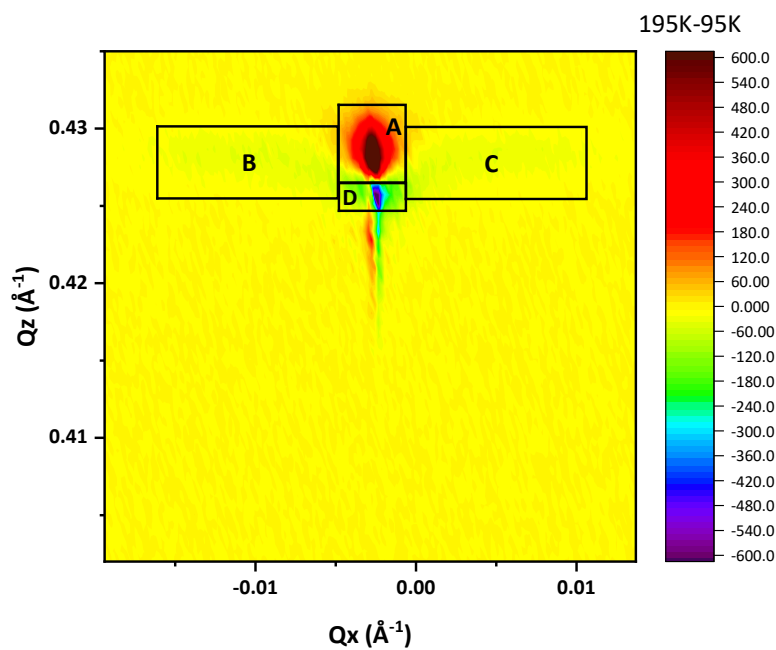

*Figure S9:* RSM intensity difference of C-cut sample between 195K and 95K

| Absolute Intensity change | 95K | 118K | 135K | 155K | 175K |
|---|---|---|---|---|---|
| Area A | 37267.0 | 29522.0 | 22232.0 | 13799.0 | 5921.0 |
| Area B | 8362.1 | 6462.8 | 5048.4 | 3106.9 | 2080.4 |
| Area C | 7847.2 | 6115.1 | 4462.7 | 2983.0 | 2061.8 |
| Area D | 10485.0 | 8083.1 | 5652.1 | 4070.0 | 1990.2 |
|  |  |  | Total Peak Intensity @ 195K | | 216030 |
| Monoclinic Phase fraction: 1.5(B+C)/total peak intensity | 11.25% | 8.73% | 6.60% | 4.23% | 2.88% |

***Table S10***: *Absolute change in peak intensities in different regions of the RSM (as indicated in Fig. S9) as a function of temperature. See text for extraction of structural phase fractions from this analysis.*

We also performed an RSM differentials analysis similar to this on the A-cut and M-cut samples. This analysis shows an excellent agreement between the SPT completion temperatures with respect to the weighted sum method described above (see **Figure S12**).

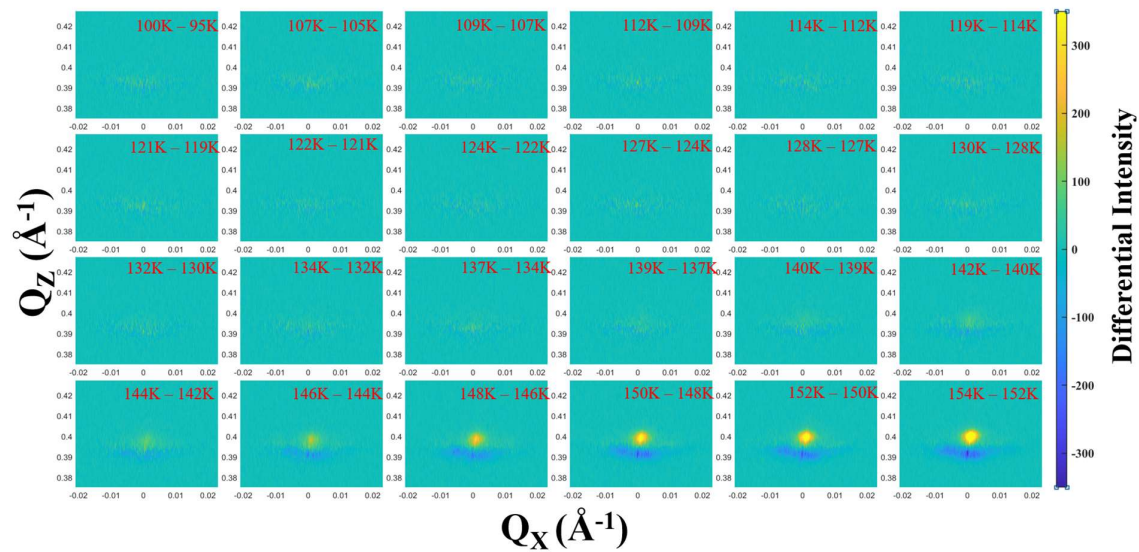

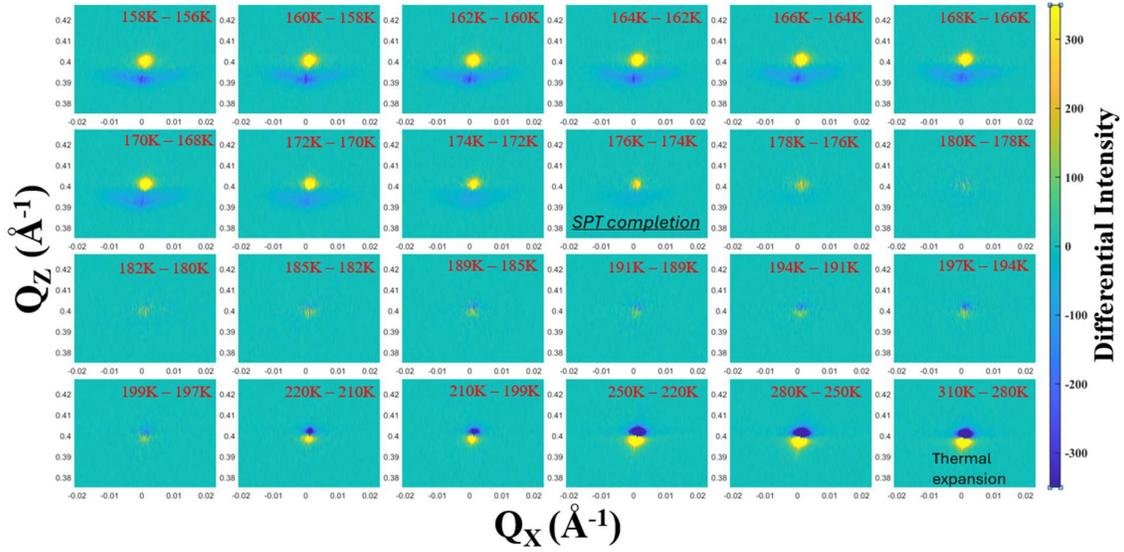

***Figure S11:*** *RSM intensity differentials acquired during heating in the A-cut sample (high temperature minus low temperature). The RSM differential vanishes at ~176 K in accordance with the RSM least-mean-squares analysis (**see Fig. S12**). At higher temperatures only one peak is observed and shifts downwards in $Q_z$ with increasing temperature due to thermal expansion.*

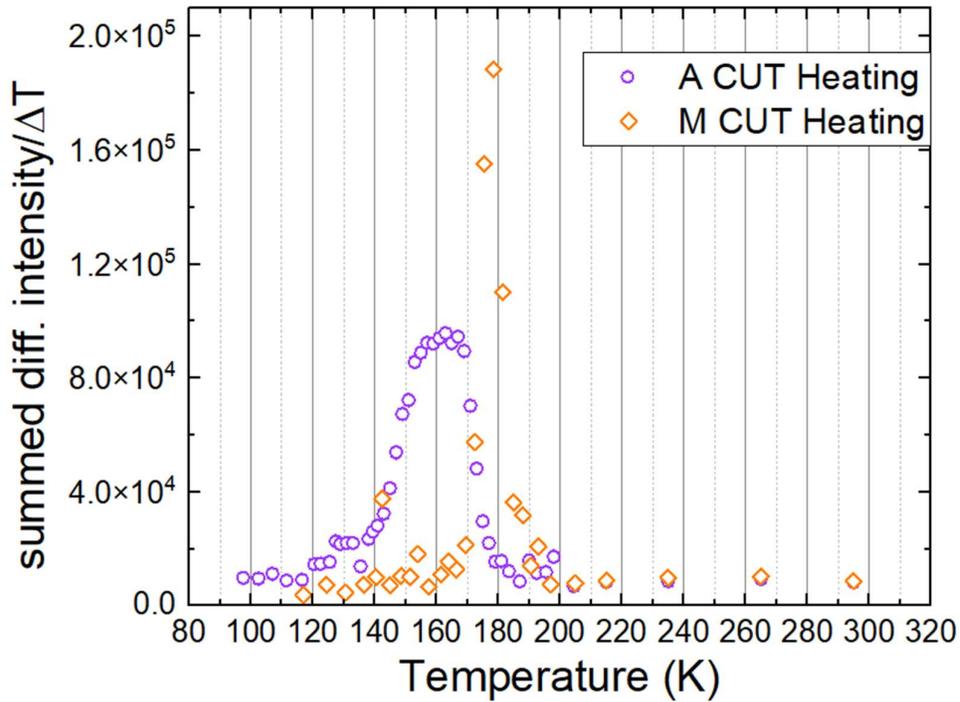

***Figure S12: RSM intensity differentials analysis for the A-cut and M-cut samples.*** *The total intensity in the differential intensities are divided by difference between the temperatures at which the two RSMs were acquired, and plotted vs their average. The temperatures at which the SPT to the rhombohedral phase is complete, agree very well with those obtained from the analysis shown in Fig. 2d&e in the main text (176 K for A-cut and 190 K for M-cut).*